\documentclass[english,preprint,showpacs,preprintnumbers,amsmath,amssymb]{revtex4}
\usepackage[T1]{fontenc}
\usepackage[latin9]{inputenc}
\usepackage{amsmath}
\usepackage{amssymb}

\makeatletter
\@ifundefined{textcolor}{}
{%
 \definecolor{BLACK}{gray}{0}
 \definecolor{WHITE}{gray}{1}
 \definecolor{RED}{rgb}{1,0,0}
 \definecolor{GREEN}{rgb}{0,1,0}
 \definecolor{BLUE}{rgb}{0,0,1}
 \definecolor{CYAN}{cmyk}{1,0,0,0}
 \definecolor{MAGENTA}{cmyk}{0,1,0,0}
 \definecolor{YELLOW}{cmyk}{0,0,1,0}
 }

\@ifundefined{textcolor}{}{%
 \definecolor{BLACK}{gray}{0}
 \definecolor{WHITE}{gray}{1}
 \definecolor{RED}{rgb}{1,0,0}
 \definecolor{GREEN}{rgb}{0,1,0}
 \definecolor{BLUE}{rgb}{0,0,1}
 \definecolor{CYAN}{cmyk}{1,0,0,0}
 \definecolor{MAGENTA}{cmyk}{0,1,0,0}
 \definecolor{YELLOW}{cmyk}{0,0,1,0}
 }

\def\be{\begin{equation}}
\def\ee{\end{equation}}

\def\bea{\begin{eqnarray}}
\def\eea{\end{eqnarray}}

\usepackage{babel}

\usepackage{babel}

\makeatother

\usepackage{babel}
\begin{document}

\title{Bijection between spin $S=\frac{p^{M}-1}{2}$ and a cluster of $M$
spins $\sigma=\frac{p-1}{2}$}

\author{N. Sh. Izmailian$^{1,2}$, Onofre Rojas$^{3}$ and S. M. de Souza$^{3}$}

\affiliation{$^{1}$ Yerevan Physics Institute, Alikhanian Br. 2, 375036 Yerevan,
Armenia}

\affiliation{$^{2}$ Institute of Physics, Academia Sinica, Nankang, Taipei 11529,
Taiwan}

\affiliation{$^{3}$ Departamento de Ciências Exatas, Universidade Federal de
Lavras, CP 3037, 37200-000 Lavras, MG, Brazil}
\begin{abstract}
We propose a general method by which a spin-$S$ is decomposed into
spins less than $S$. We have obtain the exact mapping between spin
$S=\frac{p^{M}-1}{2}$ and a cluster of $M$ spins $\sigma=\frac{p-1}{2}$.
We have discuss the possible applications of such transformations.
In particular we have show how a general $d+1$ dimensional spin-$\frac{p-1}{2}$
model with general interactions can be reduced to $d$-dimensional
spin-$S$ model with $S=\frac{p^{M}-1}{2}$.
\end{abstract}


\maketitle

\section{Introduction}

The investigation of spin-$S$ models is important for applications
and also for clarification of critical phenomena. The spin-1/2 Ising
model has been extensively investigated because it has very wide applications
to many interesting problems in different scientific areas. The general
spin-1 model with up-down symmetry was introduced by Blume, Emery
and Griffits (BEG) \cite{blume1971} for a model of $He^{3}-He^{4}$
mixtures. A spin-3/2 model was proposed by Krinsky and Mukamel \cite{krinsky1975}
for a model of ternary fluid mixtures. The spin-$S$ model with higher
spin ($S>3/2$) have been studied in much less detail because of computational
complexity increasing with spin.

So far a few exact solution have been obtained for spin-$S$ models.
In 1944 the spin-1/2 square lattice Ising model has been solved exactly
by Onsager in his seminal paper \cite{onsager1944}. All of the numerous
attempts to extend the exact solution to the systems with $S>1/2$
are failed, except for the exact results obtained for some high-spin
models than can be reduced to known solvable models, like free-fermion
models \cite{izmailian1996,izmailian1999,rojas2009}, spin-1/2 Ising
model \cite{horiguchi1986,wu1986,izmailian1994}, Ashkin-Teller model
\cite{horiguchi1995,rosenbaum2003} under certain constrained conditions,
which imposed on different coupling constants in those spin-$S$ models.
Mapping between models is an important tool for the study of exactly
solvable models. The aim of this paper is to present one of such mapping,
namely, the bijection (one-to-one transformation) between a general
spin-$S$ systems, with $S=\frac{p^{M}-1}{2}$ and a system consist
from a cluster of $M$ spins $\sigma$ with $\sigma=\frac{p-1}{2}$.

The present article is organized as follows. In section II we present
the general spin-spin transformations between spin $S=\frac{2^{M}-1}{2}$
and a cluster of $M$ spins $\sigma=\frac{p-1}{2}$. In section III
we give the expression for the general inverse spin-spin transformation.
In the next section we discuss the application of our funding. In
particular we have show that a general $d+1$ dimensional spin-$\frac{p-1}{2}$
model with general interactions can be reduced to d-dimensional spin-$S$
model with $S=\frac{p^{M}-1}{2}$ and finally in section V we give
our conclusions.

\section{A general spin-spin transformation}

Consider a system of $N$ particles with spin variable $S=(p^{M}-1)/2$
with $p$ and $M$ are a positive integer. For the $j$ ($j=1,2,...,N$)
particle, $S_{j}$ is the $z$ component of the spin operator and
its eigenvalues are $\{S,S-1,...,-S+1,-S\}$. The system has a Hamiltonian
$H(\{S_{j}\})\equiv H(S_{1},S_{2},...,S_{N})$ and the partition function
given by
\begin{equation}
Z=\sum_{\{S_{j}\}}\exp{\left[-\beta H(\{S_{j}\})\right]},
\end{equation}
 where the sum is over all $(2S+1)^{N}$ possible spin configurations.

Let us consider the following representation for the eigenvalues of
the $z$ component of the spin operator $S_{j}$
\begin{eqnarray}
S_{j} & = & \sum_{i=1}^{M}p^{i-1}\sigma_{i,j}=\sigma_{1,j}+p\,\sigma_{2,j}+...+p^{M-1}\sigma_{M,j},\label{transformation}
\end{eqnarray}
 where $\sigma_{i,j}$ are variables which takes values $\{\frac{p-1}{2},\frac{p-3}{2},...,-\frac{p-1}{2}\}$.
For even $p=2k$ we have a clusters of $M$ half integer spins $\sigma=k-1/2$
which represent the half integer spin $S=\frac{(2k)^{M}-1}{2}$ and
for odd $p=2k+1$ we have a clusters of $M$ integer spins $\sigma=k$
which represent the integer spin $S=\frac{(2k+1)^{M}-1}{2}$.

The transformation given by Eq. (\ref{transformation}) gives a bijection
(one-to-one correspondence) between spin $S=(p^{M}-1)/2$ and a cluster
of $M$ spins $\{\sigma_{1},\sigma_{2},...,\sigma_{M}\}$ where $\sigma_{i}=(p-1)/2$
for $i=1,2,...,M$. The total number of spin components of spin $S=(p^{M}-1)/2$
are
\[
2S+1=p^{M},
\]
 which is exactly the total number of spin configurations of a cluster
of $M$ spins $\sigma=(p-1)/2$. Thus we can say that the set of spin-$S$
and set of spins $\{\sigma_{1},\sigma_{2},...,\sigma_{M}\}$ are equivalent.
Note that in the case $p=2$, we obtained the condition
\[
2S+1=2^{M},
\]
 which are necessary condition to express the operator of spin-$S$
in terms of fermions \cite{dobrov2003}.

By means of Eq. (\ref{transformation}) the Hamiltonian $H(\{S_{j}\})$
may be expressed as a function of the $\sigma_{i,j}$
\[
H(\{S_{j}\})=H(\{S_{j}(\{\sigma_{i,j}\})\}).
\]
 We can therefore say that the new spin variables $\sigma_{i}$ are
independent and consequently the summation in $\sum_{\sigma_{1},\sigma_{2},...,\sigma_{M}}\exp{(-\beta H(\{\sigma_{i}\}))}$
can be carried out independently. So we may write the partition function
as the sum over all possible ($p^{MN}$) spin configurations of a
cluster of $M$ spins $\sigma_{i}=\frac{p-1}{2}$ ($i=1,2,...,M$)
instead of taking sum over $2S+1$ spin configuration of spin-$S$
\begin{equation}
Z=\sum_{\{S_{j}\}}\exp{\left[-\beta H(\{S_{j}\})\right]}=\sum_{\{\sigma_{i,j}\}}\exp{\left[-\beta H(\{S_{j}(\{\sigma_{i,j}\})\})\right]}\label{sum}
\end{equation}
 The situation is different from the case considered by Griffiths
in his paper \cite{griffiths1969}, where he introduced the mapping
between spin $S=p/2$ and a cluster of $p$ spins $\sigma=1/2$ in
the following form
\begin{equation}
S=\sum_{i=1}^{p}\sigma_{i}=\sigma_{1}+\sigma_{2}+...+\sigma_{p}.\label{Grif}
\end{equation}
 In that case the mapping is not a bijection and the Eq. (\ref{sum})
is changed to
\begin{equation}
\sum_{\{S_{j}\}}\exp{(-\beta H(\{S_{j}\}))}=\sum_{\{\sigma_{i,j}\}}\exp{(-\beta\bar{H}(\{\sigma_{i,j}\}))},\label{sum1}
\end{equation}
 where Griffiths introduced the so called weight function $W_{j}(\sigma_{1,j},\sigma_{2,j},...,\sigma_{p,j})$
in definition of the analog Hamiltonian $\bar{H}(\{\sigma_{i,j}\})$
\begin{equation}
\bar{H}(\{\sigma_{i,j}\})=H(S(\{\sigma_{i,j}\}))-\frac{1}{\beta}\sum_{j=1}^{N}\ln W_{j}(\{\sigma_{i,j}\})\label{Hbar}
\end{equation}
 where $\beta=(kT)^{-1}$ is the inverse temperature and analog Hamiltonian
is temperature dependent. In our case the weight function is equal
to 1 ($W_{j}=1$) and Eqs. (\ref{sum1}) and (\ref{Hbar}) become
Eq. (\ref{sum}).

\section{A general inverse spin-spin transformation}

Let us now consider the spin-spin transformation inverse to transformation
given by Eq. (\ref{transformation}). We want to express a spin $\sigma=\frac{p-1}{2}$
as a function of spin-$S$. The general expression of such inverse
transformation can be written in the following form
\begin{equation}
\sigma_{m}(s)=\sum_{j=0}^{2S}A_{j}(s)P_{j,m},\label{eq:sigma-coeff-s}
\end{equation}
 where
\begin{equation}
A_{j}(s)=\frac{(-1)^{2S+j}}{j!(2S-j)!}\prod_{\substack{i=0\\
i\ne j
}
}^{2S}(s+S-i),\label{Ajs}
\end{equation}
 $P_{j,m}$ are the projection of spin-$\sigma$ and $m=1,2,...,M'$
with $M'>M$, where $M'$ represents the number of permutation of
the $\{P_{1,m},\ldots,P_{p^{M},m}\}$. Let us consider as an first
example the case $p=2$, so the only possible values of $P_{j,m}$
is $\pm\frac{1}{2}$ (in reference \cite{subm-lett-11} the $P_{j,m}$
was considered as $\pm1$). For the case of spin-1 we have $p=3$,
the possible values of $P_{j,m}$ should be $\{-1,0,1\}$. Similarly
for $p=4$, the only possible values $P_{j,m}$ are $\{-\frac{3}{2},-\frac{1}{2},\frac{1}{2},\frac{3}{2}\}$.
In general the possible values of $P_{j,m}$ are $\{-S,-S+1,\ldots,S-1,S\}$.
It is clear that the transformation given by Eqs. (\ref{eq:sigma-coeff-s})
and (\ref{Ajs}) is the most general transformation which maps the
set of spins $\sigma_{m}=\frac{p-1}{2}$ ($m=1,2,...,M'$) with arbitrary
integer values of $p$ to a general spin-$S$, where $S$ can take
as integer as half-integer values.

It is also clear from Eqs. (\ref{eq:sigma-coeff-s}) and (\ref{Ajs})
that $\sigma_{m}(s)$ is a polynomial in $s$ of degree $2S$, which
we can write in the following form
\begin{equation}
\sigma_{m}(s)=\sum_{j=0}^{2S}\alpha_{m,j}s^{j},\label{eq:sigma-coeff-s1}
\end{equation}
 whose coefficient $\alpha_{m,j}$ can be defined using the result
obtained in reference \cite{rojas2009}.

Let us chose the special values of $P_{j,m}$ that we can map exactly
$M$ particles with spin $\sigma=(p-1)/2$ to spin $S=(p^{M}-1)/2$
\begin{equation}
\sigma_{m}(s)=\sum_{j=0}^{2S}\frac{(-1)^{2S+j}P_{j,m}}{j!(2S-j)!}\prod_{\substack{i=0\\
i\ne j
}
}^{2S}(s+S-i),\label{eq:sigma-coeff-s2}
\end{equation}
 where $m$ take integer values from 1 to $M$ and $P_{j,m}$ are
given by
\begin{equation}
P_{j,m}=\left[j\, p^{m-M}\right]-p\left[j\, p^{m-1-M}\right]-\frac{p-1}{2},\label{Pjm}
\end{equation}
 by $[x]$ we mean the less integer of any real $x$. With such values
of $P_{j,m}$ the Eq. (\ref{eq:sigma-coeff-s}) and (\ref{Ajs}) gives
us exactly $M$ different values of $\sigma_{m}$ ($m=1,2,...,M$)
and such transformation again give us one-to-one correspondence between
set of spins $\sigma_{m}$ ($m=1,2,...,M$) and half-integer spin-$S$
with $S=\frac{p^{M}-1}{2}$. Note, that this special values of $P_{j,m}$
already was obtained for the particular case $p=2$ in reference \cite{subm-lett-11}.
It is also worth to note that the transformation given by Eq. (\ref{eq:sigma-coeff-s})
and (\ref{Ajs}) also is a generalization of the previous result obtained
by Joseph \cite{joseph} for the case $p=2$ and for the particular
case of the projection $P_{j,m}$ ($P_{j,m}=1$ for all values of
$j$ and $m$).

\section{Applications}

In Sec. II we consider a system of $N$ particles with spin variable
$S_{j}$ ($j=1,2,...,N$) and represent the spin variables for the
$j$-th particle, $S_{j}$ in the form Eq. (\ref{transformation}),
which give us a bijection between spin $S=(p^{M}-1)/2$ and a cluster
of $M$ spins ($\sigma_{1},\sigma_{2},...,\sigma_{M}$) where $\sigma_{i}=(p-1)/2$
for $i=1,2,...,M$. In Sec. III we consider the spin-spin transformation
inverse to transformation given by Eq. (\ref{transformation}). From
the general inverse spin-spin transformation (see Eqs. (\ref{eq:sigma-coeff-s})
and (\ref{Ajs})) we chose the special values of $P_{j,m}$ and obtain
the inverse spin-spin transformations given by Eqs. (\ref{eq:sigma-coeff-s2})
and (\ref{Pjm}) that can map exactly $M$ particles with spin $\sigma=(p-1)/2$
to spin $S=(p^{M}-1)/2$ inverse to transformation system of $N$
particles with spin variable $S_{j}$ ($j=1,2,...,N$).

Let us now consider few examples:

\subsection{Spin-spin transformation for $M=2$ and $p=2$}

For the case $M=2$ and $p=2$ the transformations given by Eqs. (\ref{transformation}),
(\ref{eq:sigma-coeff-s2}) and (\ref{Pjm}) reads as
\begin{equation}
S_{j}=\sigma_{1,j}+2\sigma_{2,j}\label{SM2p2}
\end{equation}
 and $\sigma_{1,j}$ and $\sigma_{2,j}$ are given by
\begin{eqnarray}
\sigma_{1,j} & = & \tfrac{13}{12}s-\tfrac{1}{3}s^{3},\\
\sigma_{2,j} & = & -\tfrac{7}{6}s+\tfrac{2}{3}s^{3}.\label{sM2p2}
\end{eqnarray}
 The above transformation gives us bijection between spin $S=3/2$
and pair of Ising spins $\sigma=1/2$. Such transformation has been
already used by many authors \cite{izmailian1996,horiguchi1996,rosenbaum2003,rojas2009}.
For example in the paper \cite{izmailian1996} the most general spin
3/2 model with up-down symmetry was solved exactly along two lines
in the parameter space of the model with the help of transformation
given by Eqs. (\ref{SM2p2}) and (\ref{sM2p2}). In the paper \cite{horiguchi1996}
the authors used the transformation given by Eqs. (\ref{SM2p2}) and
(\ref{sM2p2}) to show that spin-3/2 model is equivalent to the two-layer
Ising model with the spin-1/2. In paper \cite{rosenbaum2003} the
new type of exact solution to the generalized Ashkin-Teller model
was found with the help of the above mentioned transformation and
in the paper \cite{rojas2009} the authors present a set of exactly
solvable models, with half-integer spin-$S$ on a square-type lattice
including the case of the spin-3/2.

\subsection{Spin-spin transformation for $M=3$ and $p=2$}

For the case $M=3$ and $p=2$ the transformations given by Eqs. (\ref{transformation}),
(\ref{eq:sigma-coeff-s2}) and (\ref{Pjm}) reads as
\begin{equation}
S_{j}=\sigma_{1,j}+2\sigma_{2,j}+4\sigma_{3,j}\label{SM3p2}
\end{equation}
 and $\sigma_{1,j}$, $\sigma_{2,j}$ and $\sigma_{3,j}$ are given
by
\begin{eqnarray}
\sigma_{1,j} & = & \tfrac{1}{252}s^{7}-\tfrac{61}{720}s^{5}+\tfrac{301}{576}s^{3}-\tfrac{30251}{26880}s,\\
\sigma_{2,j} & = & -\tfrac{1}{630}s^{7}+\tfrac{17}{360}s^{5}-\tfrac{637}{1440}s^{3}+\tfrac{14887}{13440}s,\label{sM3p2}\\
\sigma_{3,j} & = & -\tfrac{4}{315}s^{7}+\tfrac{11}{45}s^{5}-\tfrac{217}{180}s^{3}+\tfrac{2161}{1680}s.
\end{eqnarray}
 The above transformation gives us bijection between spin $S=7/2$
and three Ising spins $\sigma=1/2$. Such transformation are new and
the possible applications of that transformation can be described
as follows:

1. One can used such transformation to show that a d-dimensional general
spin - 7/2 model is equivalent to the three d-dimensional layer of
spin-1/2 model.

2. Based on such correspondence one can try to find exact solvable
cases for a general two-dimensional spin-7/2 model.

Let us consider for example a most general spin-7/2 model with up-down
symmetry on a d-dimensional lattice G, whose Hamiltonian is given
by
\begin{equation}
-\beta H(S)=\sum_{<i,j>}\left\{ \sum_{\substack{\alpha,\beta=1\\
\beta\ge\alpha\\
\alpha+\beta=even
}
}^{2S}\frac{J_{\alpha,\beta}}{2}(S_{i}^{\alpha}S_{j}^{\beta}+S_{i}^{\beta}S_{j}^{\alpha})+\sum_{a=1}^{\frac{2S-1}{2}}\frac{\gamma h_{2a}}{2}(S_{i}^{2a}+S_{j}^{2a})\right\} ,\label{HS}
\end{equation}
 with $S=7/2$, where $\beta=1/kT$ us usual and $\gamma$ is coordination
number of the lattice. Here, each spin variable $S_{i}$ is defined
at a lattice site and takes one of the following value $\{7/2,5/2,...,-7/2\}$.
The summation over $\alpha$ and $\beta$ (here $\beta$ cannot be
confused with inverse temperature) from 1 to 7 and $<i,j>$ indicates
summation over the pairs of nearest neighbor sites. We have 16 nearest
neighbor interactions terms with interaction constant $J_{\alpha,\beta}$
with $\alpha,\beta=1,2...,7$, $\beta\ge\alpha$ and $\alpha+\beta=$
even. We have also three fields $h_{2a}$ with $a=1,2,3$. Totally
we have 19 interactions constants $J_{\alpha,\beta}$ and $h_{2a}$.
Hence the Hamiltonian (\ref{HS}) represents a wide class of systems.
The free energy of the system $f_{S}$ is defined by
\begin{equation}
-\beta f_{S}=\lim_{V\to\infty}\frac{1}{V}\ln{Z},\label{freeenergy}
\end{equation}
 where $V$ is the volume of the system and $Z$ is the partition
function given by
\begin{equation}
Z=\sum_{\{S_{i}\}}\exp{\left[-\beta H(S)\right]}.
\end{equation}
 Now we apply the transformations given by Eqs. (\ref{SM3p2}) to
express $S_{i}$ in terms of three Ising spins $\sigma_{1,i},\sigma_{2,i}$
and $\sigma_{3,i}$. Then we have the following Hamiltonian from Eq.
(\ref{HS}) up to a constant
\begin{eqnarray}
-\beta H(\sigma) & = & \sum_{<i,j>}\left\{ \sum_{a=1}^{3}K_{a,a}\sigma_{a,i}\sigma_{a,j}+\sum_{\substack{a,b=1\\
b>a
}
}^{3}\frac{K_{a,b}}{2}(\sigma_{a,i}\sigma_{b,j}+\sigma_{b,i}\sigma_{a,j})+\sum_{\substack{a,b=1\\
b>a
}
}^{3}\frac{K_{b,a}}{2}(\sigma_{a,i}\sigma_{b,i}+\sigma_{a,j}\sigma_{b,j})\right.\nonumber \\
 & + & \left.\sum_{\substack{a,b,c=1\\
c\ne b\ne a\\
c>b
}
}^{3}\frac{R_{a,b,c}}{2}\sigma_{a,i}\sigma_{a,j}(\sigma_{b,i}\sigma_{c,j}+\sigma_{b,j}\sigma_{c,i})+\sum_{\substack{a,b,c=1\\
c\ne b\ne a\\
c>b
}
}^{3}\frac{R_{a,c,b}}{2}\sigma_{a,i}\sigma_{a,j}(\sigma_{b,i}\sigma_{c,i}+\sigma_{b,j}\sigma_{c,j})\right.\nonumber \\
 & + & \left.\sum_{\substack{a,b=1\\
b>a
}
}^{3}R_{a,b}\sigma_{a,i}\sigma_{a,j}\sigma_{b,i}\sigma_{b,j}+R\,\sigma_{1,i}\sigma_{2,i}\sigma_{3,i}\sigma_{1,j}\sigma_{2,j}\sigma_{3,j}\right\} \label{Hsigma}
\end{eqnarray}
 which can be considered as a Hamiltonian of the three layer of spin-$\sigma$
($\sigma=1/2$) model, where each layer is represented by d-dimensional
lattice G. We have 9 two-spin interactions given by interaction constants
$K_{a,b}$ ($a,b=1,2,3$), 9 four-spin interactions given by 6 interaction
constants $R_{a,b,c}$ ($a\ne b\ne c$) and three interaction constants
$R_{a,b}$ ($b>a$) and one six-spin interaction given by interaction
constant $R$. Totally we again have 19 interaction constants ($R,R_{a,b,c},K_{a,b}$)
(as in the case of the spin-7/2 model). Dependence of the new coefficients
$R,R_{a,b},R_{a,b,c}$ from old ones $J_{\alpha,\beta}$ and $h_{a}$
are given in Appendix (see Eqs. (\ref{K11}) - (\ref{R})). Thus we
have established one-to-one correspondence between spin-7/2 model
and three layers spin-1/2 model.

The above constructions can be easily extend to arbitrary spin-$S$
(with $S=\frac{p^{M}-1}{2}$) model to show a one-to-one correspondence
between d-dimensional spin-$S$ model and $M$ d-dimensional layers
of spin-$\frac{p-1}{2}$ model.

One can consider some particular cases of the Hamiltonian given by
Eq. (\ref{Hsigma}). For example, if we imposed the following conditions
\begin{eqnarray}
R & = & 0,\nonumber \\
R_{a,b,c} & = & 0\qquad\mbox{for}\qquad a\ne b\ne c,\nonumber \\
R_{a,b} & = & 0\qquad\mbox{for}\qquad b>a,\label{cond}\\
K_{a,b} & = & 0\qquad\mbox{for}\qquad b>a,\nonumber
\end{eqnarray}
 to cancel 6-spin interactions, four-spin interactions and some two-spin
interactions we will obtained from Eq. (\ref{Hsigma}) the following
Hamiltonian
\begin{eqnarray}
-\beta H(\sigma) & = & \sum_{<ij>}\sum_{a=1}^{3}K_{a,a}\sigma_{a,i}\sigma_{a,j}+\sum_{i}\sum_{\substack{a,b=1\\
b>a
}
}^{3}\gamma K_{b,a}\sigma_{a,i}\sigma_{b,i}.
\end{eqnarray}
 In this way, the system $H(S)$ is expressed in terms of three Ising
model of $\sigma_{1},\sigma_{2}$ and $\sigma_{3}$, each on the lattice
G coupled by glue interactions $K_{b,a}$, $b>a$.

If we imposed the additional to Eq. (\ref{cond}) conditions
\begin{eqnarray}
K_{1,1}=K_{2,2}=K_{3,3}\qquad\mbox{and}\qquad K_{2,1}=K_{3,2}=K_{3,1}
\end{eqnarray}
 we will arrive to the Ising model with periodic boundary conditions
with following Hamiltonian
\begin{eqnarray}
-\beta H(\sigma) & = & K_{1}\sum_{<ij>}\left(\sigma_{1,i}\sigma_{1,j}+\sigma_{2,i}\sigma_{2,j}+\sigma_{3,i}\sigma_{3,j}\right)+\gamma K_{2}\sum_{i}\left(\sigma_{1,i}\sigma_{2,i}+\sigma_{2,i}\sigma_{3,i}+\sigma_{1,i}\sigma_{3,i}\right)\label{Hperiod}
\end{eqnarray}
 with $K_{1}=\frac{105840}{19}J_{7,7}$ and $K_{2}=360h_{6}$. The
Hamiltonian given by Eq. (\ref{Hperiod}) is equivalent to the spin
- 7/2 Hamiltonian given by Eq. (\ref{HS}) with the following constraint
on the coupling constants
\begin{eqnarray}
J_{{2,2}} & = & J_{{2,4}}=J_{{2,6}}=J_{{4,4}}=J_{{4,6}}=J_{{6,6}}=0,\label{per1}\\
J_{{1,1}} & = & {\frac{8991341559}{389120}}\, J_{{7,7}},\qquad J_{{1,3}}=-{\frac{1424553921}{48640}}\, J_{{7,7}},\qquad J_{{1,5}}={\frac{62601021}{12160}}\, J_{{7,7}},\label{per2}\\
J_{{1,7}} & = & -{\frac{764019}{3040}}\, J_{{7,7}},\qquad J_{{3,3}}={\frac{52080091}{4864}}\, J_{{7,7}},\qquad J_{{3,5}}=-{\frac{12187819}{3040}}\, J_{{7,7}},\label{per3}\\
J_{{3,7}} & = & {\frac{30625}{152}}\, J_{{7,7}},\qquad J_{{5,5}}={\frac{30821}{80}}\, J_{{7,7}},\qquad J_{{5,7}}=-{\frac{7441}{190}}\, J_{{7,7}},\label{per4}
\end{eqnarray}
 and
\begin{equation}
h_{{2}}={\frac{259}{16}}\, h_{{6}},\qquad h_{{4}}=-{\frac{35}{4}}\, h_{{6}}.\label{hper}
\end{equation}
 If we imposed another additional to Eq. (\ref{cond}) conditions
\begin{eqnarray}
K_{1,1}=K_{2,2}=K_{3,3},\qquad K_{2,1}=K_{3,2}\qquad\mbox{and}\qquad K_{3,1}=0,
\end{eqnarray}
 we will arrive to the Ising model with free boundary conditions with
following Hamiltonian
\begin{eqnarray}
-\beta H(\sigma) & = & K_{1}\sum_{<ij>}\left(\sigma_{1,i}\sigma_{1,j}+\sigma_{2,i}\sigma_{2,j}+\sigma_{3,i}\sigma_{3,j}\right)+\gamma K_{3}\sum_{i}\left(\sigma_{1,i}\sigma_{2,i}+\sigma_{2,i}\sigma_{3,i}\right),\label{Hfree}
\end{eqnarray}
 with $K_{1}=\frac{105840}{19}J_{7,7}$ and $K_{3}=-90h_{6}$. The
Hamiltonian given by Eq. (\ref{Hfree}) is equivalent to the spin-7/2
Hamiltonian given by Eq. (\ref{HS}) with the constraints given by
Eqs. (\ref{per1}) - (\ref{per4}) and
\begin{equation}
h_{2}={\frac{1429}{16}}\, h_{6},\quad h_{4}=-{20}\, h_{6}.\label{hfree}
\end{equation}
 We can note that under conditions given by Eqs. (\ref{per1}) - (\ref{hper})
and (\ref{hfree}) the spin-$S$ Hamiltonian have two free parameters.

One can also obtained some exactly solvable case by putting the glue
interactions $K_{b,a}$ to zero. In that case we have three non interacting
Ising model, each on the d-dimensional lattice G, and in the case
$d=2$ we will obtained exact solvable case for spin - 7/2 model under
following conditions between interactions constant $J_{\alpha,\beta}$
and $h_{a}$ for original spin-7/2 model
\begin{alignat}{1}
J_{{1,1}}= & {\tfrac{6276855}{512}}\, J_{{5,7}}-{\tfrac{47629545}{1024}}\, J_{{7,7}}+{\tfrac{2557047}{1792}}\, J_{{5,5}},\label{exact1}\\
J_{{1,3}}= & -{\tfrac{2064839}{256}}\, J_{{5,7}}+{\tfrac{14979545}{256}}\, J_{{7,7}}-{\tfrac{16765}{16}}\, J_{{5,5}},\\
J_{{1,5}}= & -{\tfrac{87805}{32}}\, J_{{5,7}}-{\tfrac{2794715}{32}}\, J_{{7,7}}-{\tfrac{311}{8}}\, J_{{5,5}},\\
J_{{1,7}}= & {\tfrac{3709}{16}}\, J_{{5,7}}+{\tfrac{97205}{16}}\, J_{{7,7}}+{\tfrac{50}{7}}\, J_{{5,5}}\\
J_{{3,3}}= & {\tfrac{135485}{32}}\, J_{{5,7}}+{\tfrac{1225}{4}}\, J_{{5,5}}+{\tfrac{14984641}{256}}\, J_{{7,7}},\\
J_{{3,5}}= & -{\tfrac{3871}{16}}\, J_{{5,7}}-35\, J_{{5,5}},\\
J_{{3,7}}= & -{\tfrac{35}{2}}\, J_{{5,7}}-{\tfrac{3871}{8}}\, J_{{7,7}},
\end{alignat}

\begin{equation}
J_{{2,2}}=J_{{2,4}}=J_{{2,6}}=J_{{4,4}}=J_{{4,6}}=J_{{6,6}}=0\quad\text{{and}}\quad h_{{2}}=h_{{4}}=h_{{6}}=0.\label{exact2}
\end{equation}
 The Hamiltonian under conditions given by Eqs. (\ref{exact1}) -
(\ref{exact2}) is reduced to
\begin{eqnarray}
-\beta H(\sigma) & = & \sum_{<i,j>}\left(K_{1,1}\sigma_{1,i}\sigma_{1,j}+K_{2,2}\sigma_{2,i}\sigma_{2,j}+K_{3,3}\sigma_{3,i}\sigma_{3,j}\right)\label{Hexact}
\end{eqnarray}
 where interaction constants $K_{1,1},K_{2,2}$ and $K_{3,3}$ are
given by
\begin{alignat*}{1}
K_{1,1}= & {\tfrac{114975}{16}}\, J_{{5,7}}+{\tfrac{11432925}{64}}\, J_{{7,7}}+{\tfrac{1125}{4}}\, J_{{5,5}},\\
K_{2,2}= & 91350\, J_{{5,7}}+{\tfrac{7397775}{4}}\, J_{{7,7}}+4500\, J_{{5,5}},\\
K_{3,3}= & -88200\, J_{{5,7}}-2061675\, J_{{7,7}}-3600\, J_{{5,5}}.
\end{alignat*}
 Thus we have show that three non interacting two-dimensional exactly
solvable Ising model are equivalent to two-dimensional spin-7/2 model
under conditions given by Eqs. (\ref{exact1}) - (\ref{exact2}).

\subsection{Spin-spin transformation for $M=2$ and $p=3$}

For the case $M=2$ and $p=3$ the transformations given by Eqs. (\ref{transformation}),
(\ref{eq:sigma-coeff-s2}) and (\ref{Pjm}) reads as
\begin{equation}
S_{j}=\sigma_{1,j}+3\sigma_{2,j}
\end{equation}
 and $\sigma_{1,j}$ and $\sigma_{2,j}$ are given by
\begin{eqnarray}
\sigma_{1,j} & = & s(s^{2}-1)\left(-\tfrac{27}{560}s^{2}+\tfrac{1}{560}s^{4}+\tfrac{139}{420}\right),\nonumber \\
\sigma_{2,j} & = & s(s^{2}-9)\left(\tfrac{57}{560}s^{2}-\tfrac{3}{560}s^{4}-\tfrac{31}{140}\right),\label{M2p3}
\end{eqnarray}
 The above transformations gives us bijection between spin $S=4$
and pair of spins with $\sigma=1$. Such transformation can be used
for example to show that spin - 4 model is equivalent to the two-layer
spin-1 model.

\subsection{Spin-spin transformation for $M=3$ and $p=3$}

For the case $M=3$ and $p=3$ the transformations given by Eqs. (\ref{transformation}),
(\ref{eq:sigma-coeff-s2}) and (\ref{Pjm}) reads as
\begin{equation}
S_{j}=\sigma_{1,j}+3\sigma_{2,j}+9\sigma_{3,j}
\end{equation}
 and $\sigma_{1,j}$, $\sigma_{2,j}$ and $\sigma_{3,j}$ are given
by
\begin{eqnarray}
\sigma_{1,j} & = & \left(\tfrac{15184387919}{1918955630592000}s^{2}-\tfrac{66791923009387}{192008442802176000000}s^{4}+\tfrac{22371900997}{2643031196467200000}s^{6}-\tfrac{5057645209}{40579872915456000000}s^{8}\right.\nonumber \\
 &  & -\tfrac{17330419}{226767340800}++\tfrac{2967383}{2608706115993600000}s^{10}-\tfrac{664843}{105721247858688000000}s^{12}+\tfrac{883}{45812540738764800000}s^{14}\nonumber \\
 &  & \left.-\tfrac{1}{39836991946752000000}s^{16}\right)s(s^{2}-1)(s^{2}-2^{2})(s^{2}-3^{2})(s^{2}-4^{2}),\label{M3p3s1}\\
\sigma_{2,j} & = & \left(-\tfrac{192181663909}{624923050752000000}s^{2}+\tfrac{15276178774039}{427447366714368000000}s^{4}-\tfrac{3162180475127}{1496065783500288000000}s^{6}+\tfrac{88912189981}{1329836252000256000000}s^{8}\right.\nonumber \\
 &  & +\tfrac{148211081}{128501493120000}-\tfrac{13780223389}{11968526268002304000000}s^{10}+\tfrac{126626341}{11968526268002304000000}s^{12}-\tfrac{193309}{3989508756000768000000}s^{14}\nonumber \\
 &  & \left.+\tfrac{173}{1994754378000384000000}s^{16}\right)s(s^{2}-1)(s^{2}-8^{2})(s^{2}-9^{2})(s^{2}-10^{2}),\label{M3p3s2}\\
\sigma_{3,j} & = & \left(\tfrac{841457709}{2345390215168000}s^{2}-\tfrac{627741171441}{16417731506176000000}s^{4}+\tfrac{25815639}{14857675571200000}s^{6}-\tfrac{77436279}{1836948979712000000}s^{8}\right.\nonumber \\
 &  & -\tfrac{15097}{23796572800}++\tfrac{7713}{13121064140800000}s^{10}-\tfrac{47463}{10103219388416000000}s^{12}+\tfrac{261}{13134185204940800000}s^{14}\nonumber \\
 &  & \left.-\tfrac{9}{262683704098816000000}s^{16}\right)s(s^{2}-3^{2})(s^{2}-6^{2})(s^{2}-9^{2})(s^{2}-12^{2}).\label{M3p3s3}
\end{eqnarray}
 The above transformations gives us bijection between spin $S=13$
and a cluster of three spins with $\sigma=1$. Such transformation
can be used for example to show that spin-13 model is equivalent to
the three-layer spin-1 model.

\subsection{A $d$-dimensional Ising model mapping onto single-particle spin-$S$
model}

In 70's Joseph \cite{joseph}, discussed the single-particle mapping
onto Shrödinger exchange operators \cite{shrodinger41}, using this
approach Joseph also discussed the nature of the transition in non-linear
spin-$S$ Ising model for both cases: half-odd integer spin \cite{josepha}
and integer spin \cite{josephb}. In this section we propose some
similar approach to discuss the $d$-dimensional Ising model mapping
onto single-particle spin-$S$ model. Using the presented bijections
between spin-$S$ ($S=\frac{p^{M}-1}{2}$) and sets of $M$ spins
$(\sigma_{1},\sigma_{2},...,\sigma_{M})$ with $\sigma_{m}=\frac{p-1}{2}$
is a mapping between a $d$-dimensional Ising model and a single-particle
spin-$S$ model.

Let us consider the general $d$-dimensional Ising model given by
the following Hamiltonian.
\begin{equation}
-\beta H_{M}(s)=\sum_{<m,m'>}J_{m,m'}\sigma_{m}(s)\sigma_{m'}(s)+h\sum_{m}\sigma_{m}(s),
\end{equation}
 where $J_{m,m'}$ is the coupling term between two spins $\sigma_{m}(s)$
and $\sigma_{m'}(s)$, while $s$ is the spin-$S$ momenta taking
the values $-S,-S+1,\ldots,S-1,S$, with $S=\frac{p^{M}-1}{2}$.

Performing the transformation given by Eqs. (\ref{eq:sigma-coeff-s}),
(\ref{Ajs}) and (\ref{Pjm}) we obtain the Hamiltonian for the single-particle
spin-$S$ model
\begin{equation}
-\beta H_{M}(s)=\sum_{j=0}^{p^{M}}\sum_{j'=0}^{p^{M}}A_{j}(s)A_{j'}(s)F_{j,j'},\label{eq:sing-Hm}
\end{equation}
 where
\begin{equation}
F_{j,j'}=\sum_{<m,m'>}\left[J_{m,m'}P_{j,m}P_{j',m'}+\tfrac{2h}{\gamma}(P_{j,m}+P_{j',m'})\right],
\end{equation}
 the function $F_{j,j'}$ depends only of the lattice 'structure'
and spin coupling parameter for a given system. Note that $F_{j,j'}$
does not depend of the variable $s$.

The Hamiltonian \eqref{eq:sing-Hm} can be understand as single particle
Hamiltonian with large spin-$S$ $(S=\frac{p^{M}-1}{2})$, similar
to that discussed by Joseph \cite{joseph}. Therefore the partition
function simply becomes as the summation over spin-$S$ momenta,
\begin{equation}
\mathcal{Z}_{M}=\sum_{s=-\frac{p^{M}-1}{2}}^{\frac{p^{M}-1}{2}}\mathrm{e}^{-\beta H_{M}(s)}.
\end{equation}

This is a different way to write the partition function of d-dimensional
Ising model.

As a simple example, let us consider the spin-1/2 one-dimensional
Ising model, with periodic boundary condition $\sigma_{M+1}(s)=\sigma_{1}(s)$
and assuming uniform coupling $J$ between spins, under longitudinal
magnetic field $h$, so the Hamiltonian reads as
\begin{equation}
-\beta H_{M}(s)=\sum_{m=1}^{M}\left[J\sigma_{m}(s)\sigma_{m+1}(s)+h\sigma_{m}(s)\right],
\end{equation}
 then the function $F_{j,j'}$ for this particular case becomes
\begin{equation}
F_{j,j'}=\sum_{m=1}^{M}\left[JP_{j,m}P_{j',m+1}+\tfrac{h}{2}(P_{j,m}+P_{j',m+1})\right].
\end{equation}
 Therefore, the partition function can be expressed by
\begin{equation}
\mathcal{Z}_{M}=\sum_{s=-\frac{2^{M}-1}{2}}^{\frac{2^{M}-1}{2}}\prod_{j=0}^{2^{M}-1}\prod_{j'=0}^{2^{M}-1}\mathrm{e}^{A_{j}(s)A_{j'}(s)F_{j,j'}}.\label{eq:Zm-sum}
\end{equation}
 Using the Eq. \eqref{eq:Zm-sum}, we are able to write the partition
function for fixed $M$, then the first few terms are given by
\begin{alignat}{1}
\mathcal{Z}_{2}= & {\rm e}^{\frac{J}{2}+h}+2{\rm e}^{-\frac{J}{2}}+{\rm e}^{\frac{J}{2}-h},\label{eq:Z2}\\
\mathcal{Z}_{3}= & {\rm e}^{\frac{3J}{4}+\frac{3h}{2}}+3{\rm e}^{-\frac{J}{4}+\frac{h}{2}}+3{\rm e}^{-\frac{J}{4}-\frac{h}{2}}+{\rm e}^{\frac{3J}{4}-\frac{3h}{2}},\\
\mathcal{Z}_{4}= & 4+{\rm e}^{J+2h}+4{\rm e}^{h}+2{\rm e}^{-J}+4{\rm e}^{-h}+{\rm e}^{J-2h},\\
\mathcal{Z}_{5}= & 5{\rm e}^{\frac{J}{4}+\frac{h}{2}}+{\rm e}^{\frac{5J}{4}-\frac{5h}{2}}+5{\rm e}^{-\frac{3J}{4}-\frac{h}{2}}+5{\rm e}^{-\frac{3J}{4}+\frac{h}{2}}+5{\rm e}^{\frac{J}{4}-\frac{h}{2}}+5{\rm e}^{\frac{J}{4}-\frac{3h}{2}}+{\rm e}^{\frac{5J}{4}+\frac{5h}{2}}+5{\rm e}^{\frac{J}{4}+\frac{3h}{2}}.\label{eq:Z5}
\end{alignat}
 In order to write the previous results in analogy to the solution
obtained via transfer matrix. We define some basic quantities conveniently,
such that the factors involving the largest (lowest) spin momenta
are given by
\begin{alignat}{1}
a_{M}^{\pm}= & \exp\left(-\tfrac{\beta}{M}H_{M}(\pm\tfrac{2^{M}-1}{2})\right)=\mathrm{e}^{\frac{J}{4}\pm\frac{h}{2}}.\label{eq:sing-Z-coeff}
\end{alignat}
 Similarly, we define also define the factor for next largest (lowest)
spin momenta, which reads as
\begin{equation}
c_{M}^{\pm}=\exp\left(-\tfrac{\beta}{M}H_{M}(\pm\tfrac{2^{M}-3}{2})\right)=\mathrm{e}^{\frac{M-4}{4M}J\pm\frac{M-2}{2M}h},
\end{equation}
 the gap energy of spin-$S$ between the largest (lowest) and next
largest (lowest) energy could be defined as
\begin{equation}
\Delta E^{\pm}=-\beta\left(H_{M}(\pm\tfrac{2^{M}-1}{2})-H_{M}(\pm\tfrac{2^{M}-3}{2})\right)=J\pm h.
\end{equation}
 Let us define the following factor involving the energy gap of spin-$S$,
\begin{equation}
b_{M}^{\pm}=\exp\left(\Delta E^{\pm}/4\right)=\mathrm{e}^{\frac{J\pm h}{4}}.
\end{equation}
 Using the previous definition, and by using some algebraic manipulation,
we can express the Eq. \eqref{eq:Zm-sum} alternatively as follow,
\begin{equation}
\mathcal{Z}_{M}=\left(\tfrac{a_{M}^{+}+a_{M}^{-}+\sqrt{\left(a_{M}^{+}-a_{M}^{-}\right)^{2}+4b_{M}^{+}b_{M}^{-}}}{2}\right)^{M}+\left(\tfrac{a_{M}^{+}+a_{M}^{-}-\sqrt{\left(a_{M}^{+}-a_{M}^{-}\right)^{2}+4b_{M}^{+}b_{M}^{-}}}{2}\right)^{M}.\label{eq:Z-sr-trns-M}
\end{equation}
 Using the Eq. \eqref{eq:Z-sr-trns-M}, we can verify the results
obtained in \eqref{eq:Z2} - \eqref{eq:Z5}, and higher order terms
(not shown here). It is interesting to note that $a_{M}^{\pm}$ and
$b_{M}^{\pm}$ is always independent of $M$.

Therefore the free energy in thermodynamic limit is given by
\begin{equation}
-\beta f=\lim_{M\rightarrow\infty}\ln\left(\mathcal{Z}_{M}\right)=\ln\left(\mathrm{e}^{J/4}\cosh(h/2)+\sqrt{\mathrm{e}^{J/2}\sinh(h/2)+\mathrm{e}^{-J/2}}\right).\label{Hone}
\end{equation}
 We have arrived to this results in some similar way as discussed
by Joseph \cite{joseph}. Note, that the Eq. (\ref{Hone}) can be
obtained using the transfer matrix method.

\section{Conclusion}

We have present the general spin-spin transformations between spin
$S=\frac{2^{M}-1}{2}$ and a cluster of $M$ spins $\sigma=\frac{p-1}{2}$
as well as the general inverse spin-spin transformation. We have discuss
the application of our funding. In particular we have show one-to-one
correspondence between a general spin-7/2 model on a d-dimensional
lattice G and a three Ising model, each on the lattice G coupled by
glue interactions. That results can be easily to extend to more general
case, namely, that a general d+1 dimensional spin-$\frac{p-1}{2}$
model can be reduced to d-dimensional spin-$S$ model with $S=\frac{p^{M}-1}{2}$.
The representation of particles with the spin-$S$ in terms of spins
less than $S$ seems to be very useful tool. We wish to clarify the
critical properties of spin-$S$ models with spin greater than 1/2
by using the present decomposition method in the future.

One of us (N.Sh.I) is supported by FAPEMIG (BPV-00061-10), while O.R.
and S.M. de Souza thanks FAPEMIG and CNPq for partial financial support.

\section{Appendix}

Here we give the dependence of the coefficients $K_{a,b},R_{a,b},R_{a,b,c}$
and $R$ from $J_{\alpha,\beta}$ and $h_{2},h_{4},h_{6}$
\begin{alignat}{1}
K_{1,1}= & J_{{1,1}}+{\tfrac{61}{4}}\, J_{{1,3}}+{\tfrac{3481}{16}}\, J_{{1,5}}+{\tfrac{186901}{64}}\, J_{{1,7}}+{\tfrac{3721}{16}}\, J_{{3,3}}+{\tfrac{212341}{64}}\, J_{{3,5}}\label{K11}\\
 & +{\tfrac{11400961}{256}}\, J_{{3,7}}+{\tfrac{12117361}{256}}\, J_{{5,5}}+{\tfrac{650602381}{1024}}\, J_{{5,7}}+{\tfrac{34931983801}{4096}}\, J_{{7,7}}\nonumber \\
K_{1,2}= & 2\, J_{{1,1}}+29\, J_{{1,3}}+{\tfrac{2971}{8}}\, J_{{1,5}}+{\tfrac{37417}{8}}\, J_{{1,7}}+{\tfrac{3355}{8}}\, J_{{3,3}}+{\tfrac{42697}{8}}\, J_{{3,5}}\\
 & +{\tfrac{8569045}{128}}\, J_{{3,7}}+{\tfrac{8566741}{128}}\, J_{{5,5}}+{\tfrac{212837399}{256}}\, J_{{5,7}}+{\tfrac{21014213935}{2048}}\, J_{{7,7}}\nonumber \\
K_{1,3}= & 4\, J_{{1,1}}+46\, J_{{1,3}}+{\tfrac{2371}{4}}\, J_{{1,5}}+7606\, J_{{1,7}}+{\tfrac{1891}{4}}\, J_{{3,3}}+5776\, J_{{3,5}}\\
 & +{\tfrac{4619941}{64}}\, J_{{3,7}}+{\tfrac{4389541}{64}}\, J_{{5,5}}+{\tfrac{108081833}{128}}\, J_{{5,7}}+{\tfrac{10558224391}{1024}}\, J_{{7,7}}\nonumber \\
K_{2,2}= & 4\, J_{{1,1}}+55\, J_{{1,3}}+{\tfrac{2461}{4}}\, J_{{1,5}}+{\tfrac{112435}{16}}\, J_{{1,7}}+{\tfrac{3025}{4}}\, J_{{3,3}}+{\tfrac{135355}{16}}\, J_{{3,5}}\\
 & +{\tfrac{6183925}{64}}\, J_{{3,7}}+{\tfrac{6056521}{64}}\, J_{{5,5}}+{\tfrac{276702535}{256}}\, J_{{5,7}}+{\tfrac{12641629225}{1024}}\, J_{{7,7}}\nonumber \\
K_{2,3}= & 8\, J_{{1,1}}+86\, J_{{1,3}}+{\tfrac{1861}{2}}\, J_{{1,5}}+{\tfrac{84463}{8}}\, J_{{1,7}}+{\tfrac{1705}{2}}\, J_{{3,3}}+{\tfrac{72823}{8}}\, J_{{3,5}}\\
 & +{\tfrac{3296245}{32}}\, J_{{3,7}}+{\tfrac{3103321}{32}}\, J_{{5,5}}+{\tfrac{140402443}{128}}\, J_{{5,7}}+{\tfrac{6351565585}{512}}\, J_{{7,7}}\nonumber \\
K_{3,3}= & 16\, J_{{1,1}}+124\, J_{{1,3}}+1261\, J_{{1,5}}+{\tfrac{56491}{4}}\, J_{{1,7}}+961\, J_{{3,3}}+{\tfrac{39091}{4}}\, J_{{3,5}}\\
 & +{\tfrac{1751221}{16}}\, J_{{3,7}}+{\tfrac{1590121}{16}}\, J_{{5,5}}+{\tfrac{71235151}{64}}\, J_{{5,7}}+{\tfrac{3191233081}{256}}\, J_{{7,7}}\nonumber \\
K_{2,1}= & 21\, J_{{2,2}}+{\tfrac{3003}{8}}\, J_{{2,4}}+{\tfrac{41613}{8}}\, J_{{2,6}}+{\tfrac{41181}{8}}\, J_{{4,4}}+{\tfrac{8470293}{128}}\, J_{{4,6}}\\
 & +{\tfrac{212094831}{256}}\, J_{{6,6}}+2\,\gamma\, h_{{2}}+53\,\gamma\, h_{{4}}+{\tfrac{6331}{8}}\,\gamma\, h_{{6}}\nonumber \\
K_{3,1}= & 42\, J_{{2,2}}+{\tfrac{1995}{4}}\, J_{{2,4}}+{\tfrac{25233}{4}}\, J_{{2,6}}+{\tfrac{22533}{4}}\, J_{{4,4}}+{\tfrac{4438005}{64}}\, J_{{4,6}}\\
 & +{\tfrac{107571711}{128}}\, J_{{6,6}}+4\,\gamma\, h_{{2}}+58\,\gamma\, h_{{4}}+{\tfrac{3211}{4}}\,\gamma\, h_{{6}}\nonumber \\
K_{3,2}= & 84\, J_{{2,2}}+{\tfrac{1743}{2}}\, J_{{2,4}}+9624\, J_{{2,6}}+{\tfrac{17871}{2}}\, J_{{4,4}}+{\tfrac{3150213}{32}}\, J_{{4,6}}\\
 & +{\tfrac{69380571}{64}}\, J_{{6,6}}+8\,\gamma\, h_{{2}}+92\,\gamma\, h_{{4}}+{\tfrac{2071}{2}}\,\gamma\, h_{{6}}\nonumber
\end{alignat}

\begin{alignat}{1}
R_{1,2}= & 16\, J_{{2,2}}+424\, J_{{2,4}}+6331\, J_{{2,6}}+11236\, J_{{4,4}}+{\tfrac{335543}{2}}\, J_{{4,6}}+{\tfrac{40081561}{16}}\, J_{{6,6}}\\
R_{1,3}= & 64\, J_{{2,2}}+928\, J_{{2,4}}+12844\, J_{{2,6}}+13456\, J_{{4,4}}+186238\, J_{{4,6}}+{\tfrac{10310521}{4}}\, J_{{6,6}}\\
R_{2,3}= & 256\, J_{{2,2}}+2944\, J_{{2,4}}+33136\, J_{{2,6}}+33856\, J_{{4,4}}+381064\, J_{{4,6}}+4289041\, J_{{6,6}}\\
R_{1,2,3}= & 32\, J_{{2,2}}+656\, J_{{2,4}}+9542\, J_{{2,6}}+12296\, J_{{4,4}}+176891\, J_{{4,6}}+{\tfrac{20328841}{8}}\, J_{{6,6}}\\
R_{2,1,3}= & 64\, J_{{2,2}}+1216\, J_{{2,4}}+16804\, J_{{2,6}}+19504\, J_{{4,4}}+255376\, J_{{4,6}}+{\tfrac{13111501}{4}}\, J_{{6,6}}\\
R_{3,1,2}= & 128\, J_{{2,2}}+1664\, J_{{2,4}}+21128\, J_{{2,6}}+21344\, J_{{4,4}}+267824\, J_{{4,6}}+{\tfrac{6649981}{2}}\, J_{{6,6}}\\
R_{1,3,2}= & 24\, J_{{1,3}}+420\, J_{{1,5}}+{\tfrac{11613}{2}}\, J_{{1,7}}+732\, J_{{3,3}}+{\tfrac{23253}{2}}\, J_{{3,5}}+158637\, J_{{3,7}}\nonumber \\
 & +{\tfrac{365505}{2}}\, J_{{5,5}}+{\tfrac{79674063}{32}}\, J_{{5,7}}+{\tfrac{2170481313}{64}}\, J_{{7,7}}\\
R_{2,3,1}= & 48\, J_{{1,3}}+840\, J_{{1,5}}+11613\, J_{{1,7}}+1320\, J_{{3,3}}+18933\, J_{{3,5}}+244005\, J_{{3,7}}\nonumber \\
 & +258405\, J_{{5,5}}+{\tfrac{52190943}{16}}\, J_{{5,7}}+{\tfrac{1305707655}{32}}\, J_{{7,7}}\\
R_{3,2,1}= & 96\, J_{{1,3}}+1680\, J_{{1,5}}+23226\, J_{{1,7}}+1488\, J_{{3,3}}+20586\, J_{{3,5}}+264738\, J_{{3,7}}\nonumber \\
 & +264810\, J_{{5,5}}+{\tfrac{26507103}{8}}\, J_{{5,7}}+{\tfrac{656029983}{16}}\, J_{{7,7}}
\end{alignat}

\begin{equation}
R=9\,\left(256\, J_{{3,3}}+4480\, J_{{5,3}}+78400\, J_{{5,5}}+61936\, J_{{7,3}}+1083880\, J_{{7,5}}+14984641\, J_{{7,7}}\right)\label{R}
\end{equation}

\end{document}